\documentclass[prl,aps,twocolumn,superscriptaddress,preprintnumbers,showpacs,floatfix,amsmath,amssymb]{revtex4}

\usepackage{graphicx}
\usepackage{dcolumn}
\usepackage{bm}

\usepackage{amsmath}
\usepackage{graphicx}
\usepackage{amssymb}
\usepackage{mathrsfs}
\usepackage{bbm}
\usepackage{epsfig}
\usepackage{makecell}

\newcommand{\be}{\begin{eqnarray}}
\newcommand{\ee}{\end{eqnarray}}

\begin{document}

%
\title{Towards multistage modelling of  protein dynamics    \\ 
          with monomeric Myc oncoprotein as an example
}

\author{Jiaojiao Liu} 
\email{ljjhappy1207@163.com}
\affiliation{School of Physics, Beijing Institute of Technology, Beijing 100081, P.R. China}
\author{Jin Dai}
\email{daijing491@gmail.com}
\affiliation{School of Physics, Beijing Institute of Technology, Beijing 100081, P.R. China}
\author{Jianfeng He}
\email{hjf@bit.edu.cn}
\affiliation{School of Physics, Beijing Institute of Technology, Beijing 100081, P.R. China}
\author{Antti J. Niemi}
\email{Antti.Niemi@physics.uu.se}
\affiliation{Nordita, Stockholm University, Roslagstullsbacken 23, SE-106 91 Stockholm, Sweden}
\affiliation{Department of Physics and Astronomy, Uppsala University,
P.O. Box 803, S-75108, Uppsala, Sweden}
\affiliation{Laboratoire de Mathematiques et Physique Theorique
CNRS UMR 6083, F\'ed\'eration Denis Poisson, Universit\'e de Tours,
Parc de Grandmont, F37200, Tours, France}
\affiliation{School of Physics, Beijing Institute of Technology, Beijing 100081, P.R. China}
\homepage{http://www.folding-protein.org}
\author{Nevena Ilieva}
\email{nilieval@mail.cern.ch}
\affiliation{Institute of Information and Communication Technologies, Bulgarian Academy of Sciences, 25A, Acad. G. Bonchev Str., Sofia 1113, Bulgaria}

\begin{abstract}
We propose to combine a mean field approach with all atom molecular dynamics (MD), into a
multistage  algorithm that can model protein folding and dynamics over very long time periods yet
with atomic level precision. As an example we investigate an isolated monomeric Myc 
oncoprotein that has been implicated in carcinomas including those in colon, breast and  lungs. 
Under physiological conditions a monomeric Myc is presumed to be an example of intrinsically disordered proteins,
that pose a serious challenge to existing modelling techniques. We argue
that a room temperature monomeric Myc is in a dynamical state, it oscillates between different conformations that we identify.
For this we adopt the C$\alpha$ backbone of Myc in a crystallographic heteromer as an initial
{\it Ansatz} for the monomeric structure.  We construct a multisoliton of the pertinent 
Landau free energy, to describe 
the C$\alpha$ profile with ultra high precision. We use Glauber dynamics to resolve how the multisoliton responds to repeated 
increases and decreases in ambient temperature. We confirm that the initial structure is unstable in isolation. We reveal 
a highly degenerate  ground state landscape, an attractive set towards which Glauber dynamics converges  in the limit 
of vanishing ambient temperature. We analyse the thermal stability of  this Glauber attractor using room temperature
molecular dynamics.  We identify and scrutinise a particularly stable subset in which 
the two helical segments of the original multisoliton  align in parallel, next to each other.  During the MD
time evolution of a representative structure from this subset, we observe intermittent quasiparticle 
oscillations along the C-terminal $\alpha$-helix, some of which resemble a translating Davydov's Amide-I soliton. We propose that the 
presence of oscillatory motion is in line  with the expected intrinsically disordered character of Myc.
\end{abstract}

%

\pacs{87.15.Cc, 82.35.Lr, 36.20.Ey
}


\maketitle

\section{Introduction}

All atom molecular dynamics (MD) \cite{Rapaport-2004} aims to simulate the time evolution of  every single atom in a given protein, including solvent \cite{Gelman-2015}. 
It produces a discrete and  piecewise linear 
time trajectory of each atom, as a  solution of a discretised  (semi)classical Newton's equation. 
Thus the dimensionless ratio between the
iteration time step $\Delta t$ and the  time scale $\tau$ of a characteristic atomic motion 
\begin{equation}
e \sim \frac{\Delta t}{\tau}
\label{e}
\end{equation}
should be small. Usually $\tau$ relates to the frequency of a covalent bond oscillation that has a duration of a few femtoseconds.
As a result  $\Delta t$ should be very short  and canonical 
values are around 1-2 femtoseconds. The need for such a short time step makes an all atom approach to 
protein dynamics an extreme computational challenge \cite{Gelman-2015,Shaw-2008,Shaw-2009}. For example, the folding time of a myoglobin
is around 2.5 seconds \cite{Jennings-1993} which can be considered as a fairly representative duration in 
the case of many proteins.  At the same time MD  
can at best produce around ten microseconds  of {\it in vitro} folding trajectory per  day {\it in silico}
\cite{Shaw-2012}, and this in the case of proteins which are much shorter than myoglobin.
It would probably take close to a thousand years for presently available computers, to simulate a 
single all atom folding trajectory of myoglobin. Moreover,  the currently available all atom force fields are not perfect \cite{Gelman-2015}. 
Their limitations tend to essentially affect a folding trajectory 
no later than around ten microseconds 
\cite{Shaw-2012}. Coarse-grained 
techniques are being developed to overcome the bottle-neck of short time 
steps, but with loss in  accuracy \cite{Gelman-2015,Saunders-2013}. 

\begin{figure}[h]         
\centering            
  \resizebox{8 cm}{!}{\includegraphics[]{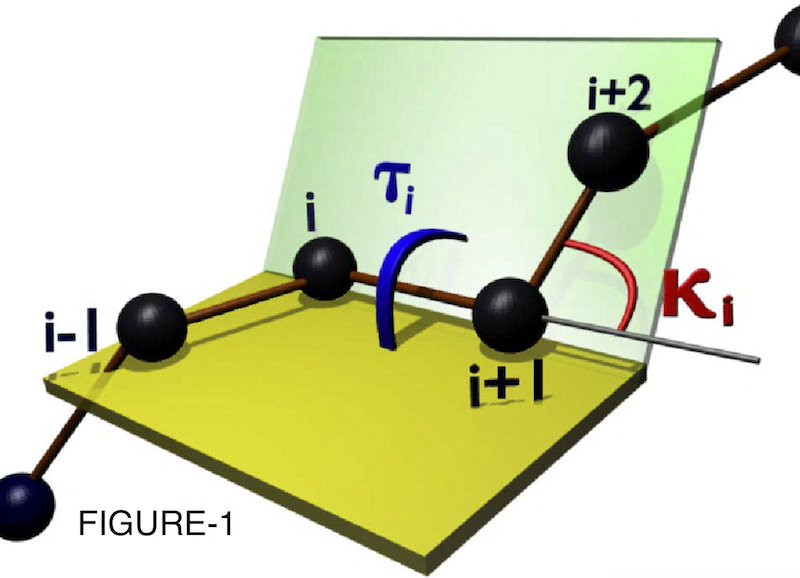}}
\caption {\small  {\it Color online:} Definition of bond ($\kappa_i$) and torsion ($\tau_i$) angles in relation to the $i^{th}$ C$\alpha$ 
atom.}   
\label{fig-1}    
\end{figure}
\begin{figure}[h]         
\centering            
  \resizebox{8 cm}{!}{\includegraphics[]{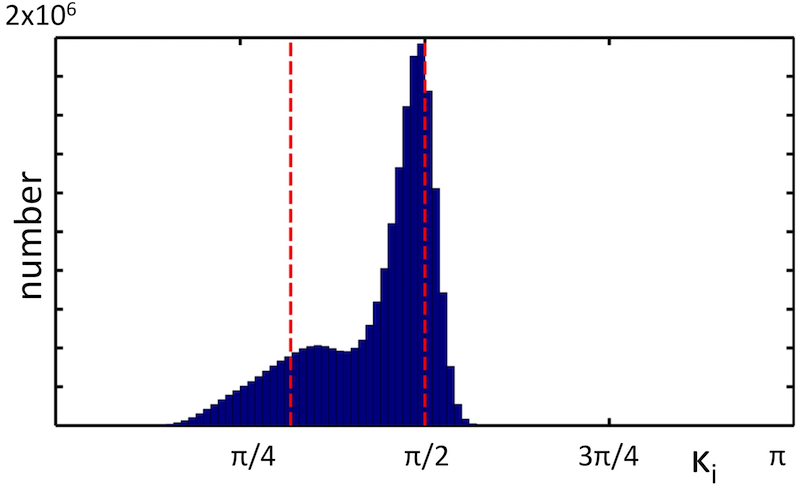}}
\caption {\small  {\it Color online:} Distribution of bond angles $\kappa$ in crystallographic PDB structures. Note that for $\alpha$-helices $\kappa \approx \pi/2$ and for $\beta$-strands $\kappa \approx 1$.}   
\label{fig-2}    
\end{figure}
\begin{figure}[h]         
\centering            
\vspace{0.2cm}  \resizebox{8 cm}{!}{\includegraphics[]{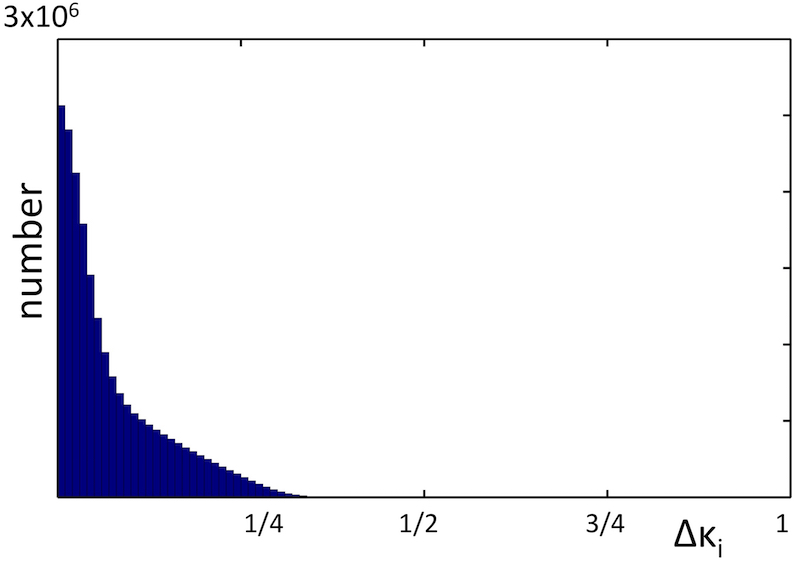}}
\caption {\small  {\it Color online:} The distribution of values (\ref{kappadiff}) along the C$\alpha$ backbone during
the Villin simulation in Anton \cite{Lindorff-2011}. Based on data obtained from authors.}   
\label{fig-3}    
\end{figure}

There are  many examples in Physics, where a  description  in terms of fundamental level constituents 
is too strenuous. In such cases the concept of a mean field theory can provide a pragmatic alternative 
\cite{Goldenfeld-1992}.   It has been proposed that a  mean field approach could be  introduced  to model proteins
in terms of the  C$\alpha$ backbone \cite{Danielsson-2010,Chernodub-2010,Molkenthin-2011,Krokhotin-2012a,Krokhotin-2012b,krokhotin-2014,Sieradzan-2014,fadde,les-houches}. For this we note that 
any biologically relevant time scale is long in comparison to the period of a
covalent bond oscillation. Thus the distance between two neighbouring C$\alpha$ atoms can be approximated by 
the average value which is around 3.8 \AA ngstr\"om.
A Landau free energy then engages only the bond angles $\kappa\in [0,\pi)$ and 
the torsion  angles
$\tau\in [-\pi, \pi)$ of the C$\alpha$ skeletal as structural order parameters, as shown in Figure \ref{fig-1}. 
%
%
%
%
%
%
%
%
%
%
%
%
%
%
%
%
Moreover, the bond angles are known to have very small variations along a protein backbone,  both in static Protein Data Bank
(PDB) \cite{Berman-2000}  structures and
during dynamical MD simulations, as confirmed by Figures \ref{fig-2} and \ref{fig-3}.  
%
%
%
%
%
%
%
%
%
%
%
%
%
%
%
%
%
%
%
%
%
%
%
%
%
%
%
%
%
Thus, the relative difference in values of $\kappa_i$ between two neighbouring residues 
\begin{equation}
\Delta \kappa_i = \frac{|\kappa_{i+1} - \kappa_i|}{\pi}
\label{kappadiff}
\end{equation}
is small and  can be employed 
as an expansion parameter in lieu of (\ref{e}). In particular, since an expansion in (\ref{kappadiff}) 
does not relate to any time scale, a mean field description 
can at least in principle describe time trajectories over any  time period.  

Here we propose to combine mean field theory with all atom molecular dynamics into
a multistage algorithm to model protein dynamics, and in particular intrinsically unstructured proteins,
over long time periods yet with atomic level scrutiny. 
We use mean field theory 
to leap over high energy barriers and  long time periods. 
We refine the structure to atomic level by following how it
evolves over a short time period using all atom MD.  
The algorithm goes as follows. 
We first construct  the Landau free energy that models a given protein structure. The initial {\it Ansatz} can either
be taken from PDB, or it can be constructed using  homology modelling with all atom MD refinement as in
\cite{Shaw-2012}. We then proceed to  
locate the  minimum energy 
configurations of the Landau free energy. For this we use
the Glauber algorithm \cite{Glauber-1963,Berg-2004} to 
repeatedly  increase and decrease the ambient temperature
between very high and very low values.  In the limit of very low temperatures, at the end of heating and cooling, 
the structure settles near a local minimum of the free energy. Thus, by numerously repeating a heating and  
cooling cycle  we can reveal the  low energy landscape,  as the set of structures 
towards which the C$\alpha$ backbone becomes attracted in the low temperature limit of 
the Glauber algorithm. Once we have found such a Glauber attractor of low energy structures, 
we proceed to  refine and scrutinise it using all atom MD. 
When certain pre-determined convergence and stability criteria are met,  
the simulation is considered complete and the algorithm is terminated. Otherwise, the procedure 
is repeated.

\begin{figure}[h]         
\centering            
\vspace{0.2cm}  \resizebox{8 cm}{!}{\includegraphics[]{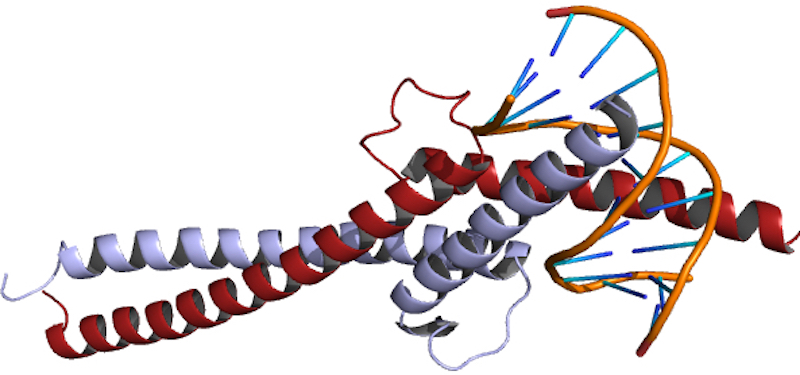}}
\caption {\small  {\it Color online:} The C$\alpha$ backbone of the crystallographic PDB structure 1NKP, with a segment of DNA. 
Myc in red, Max in blue. }   
\label{fig-4}    
\end{figure}

As an example we investigate the topography of Glauber attractor in the case of the biomedically highly important
Myc proto-oncogene protein \cite{Sheiness-1978,DePinho-1987,Meyer-2008,Arrowsmith-2012,Dang-2012,Nair-2003,Follis-2009}. 
Malfunctioning and over-expression of Myc has been implicated in a number of human cancers, from lymphomas and leukemias to
carcinomas  in colon, breast and  lungs. Thus, Myc is considered a 
promising target for cancer therapeutics and development of anti-cancer drugs.
Under physiological conditions a monomeric Myc is presumed to be 
intrinsically unstructured and it is not known to have any direct biological effect. 
Apparently, Myc becomes functionally active only when it stabilises into a DNA binding basic-helix-loop-helix-leucine-zipper 
conformation upon heterodimerization with Max; see Figure \ref{fig-4}.
%
%
%
%
%
%
%
%
%
%
%
%
%
%
%
%
%
As a component of the heterodimer, Myc then participates in processes such as
cell cycle progression, apoptosis and cellular transformation, by regulating the transcription of  the relevant 
target gene.

We analyse a monomeric Myc in isolation, as 
a biomedically  important exemplar to develop
our methodology: There should be a correlation between the conformation 
of a monomeric Myc and the rate at which it can heterodimerize with Max {\it in vivo}. 
We start our analysis from a crystallographic structure of 
the Myc-Max heteromer which is bound to DNA; we use the structure with PDB code 
1NKP  \cite{Nair-2003}, which is shown in Figure \ref{fig-4}. 
We construct the initial Landau free energy {\it Ansatz} of monomeric
Myc using the conformation it has
in the heterodimer. We subject the ensuing multisoliton model of Myc to repeated heating 
and cooling cycles using the Glauber algorithm. 
We categorize the C$\alpha$ structures towards which it becomes attracted upon cooling. We observe
that the resulting Glauber attractor accumulates along a linear trajectory 
in terms of root-mean-square distance (RMSD) and radius of gyration $R_g$. This trajectory emanates from
the initial multisoliton structure and proceeds  towards decreasing $R_g$, and energy. We identify
five different structural clusters along the mean field trajectory, and we select a representative from each cluster 
for MD simulation stability analysis. We use molecular dynamics package GROMACS 4.6.3 \cite{Hess-2008}
with the united-atom force field GROMOS53a6, which we have previously analysed and 
compared with all-atom force fields 
CHARMM27 and OPLS/AA \cite{Dai-2016b} in a closely related context; we deduce that GROMOS53a6 is the most reliable among
the three force fields, for the present purposes.
We perform the MD simulation near 
room temperature at 290 K and we limit the simulation duration  to 50 nanoseconds {\it in vitro}: Since we are interested in 
the local stability and refinement of the initial structure, we do not attempt a full scale all atom MD search of a 
folded Myc. Besides, we doubt that a any presently available computer power is sufficient for such an analysis.
We find that in four of the clusters we identify,
the MD trajectory drifts away from the cluster. Accordingly, these clusters are unstable 
under MD time evolution. However, the fifth cluster is remarkably stable under MD evolution. The MD simulation is only   
slightly re-adjusting the positioning of the backbone and side chain atoms. But during the MD evolution of the representative 
that we have chosen from the apparently stable cluster, we also  observe intermittent oscillatory behaviour, some aspects of which 
resemble a propagating, asymmetric Amide-I Davydov soliton \cite{Davydov-1973,Scott-1992}.


Our Glauber dynamics simulations are swift. A full heating and cooling cycle takes only
around a minute {\it in silico}, when we use a single processor in a current Apple Pro desktop computer. 
The MD analysis is much more tedious: The ideal length of a  MD trajectory seems to be 
around a few microseconds \cite{Shaw-2012}.  However, the computer resources that are available to us  in practice 
limit the duration of our MD trajectories to around a hundred or so nanoseconds. 
Moreover, we do not repeat our  multiscale algorithm  
beyond its first level of iteration. Since we identify an apparently MD stable assembly 
already after a single iteration,  the 
example we present serves as a  proof-of-concept exercise.  For a firmer conclusion
on Myc and its landscape of assemblies the technology described in \cite{Shaw-2012} should be used.

\section{Methods}

\subsection{Mean field theory}
A mean field model of a protein is built as follows:  Most biologically relevant processed have a time scale which is
very long in comparison to the period of a covalent bond oscillation. Thus, over any
biologically relevant time period we can approximate the distance between two neighbouring C$\alpha$ atoms  
with the average value 3.8 \AA ~ of a crystallographic PDB structure. 
The skeletal C$\alpha$ bond $\kappa$ and torsion $\tau$ angles that we
define in Figure \ref{fig-1}, then  constitute a complete set of structural order parameters. 
As shown in Figures \ref{fig-2} and \ref{fig-3}, 
the bond angles are relatively rigid and slowly varying; the differences $\Delta \kappa_i$ in (\ref{kappadiff}) 
are small. Thus, the Landau free energy $E(\kappa,\tau)$ can be expanded in
powers of these differences. 
A detailed analysis  in ~\cite{Danielsson-2010,Chernodub-2010,Molkenthin-2011,Krokhotin-2012a,Krokhotin-2012b,krokhotin-2014,Sieradzan-2014,fadde,les-houches} shows that in the limit of small $\Delta \kappa_i$  
the free energy admits the following expansion
\[
 E(\kappa,\tau)  = 
\sum\limits_{i=1}^{N-1}  
\Delta \kappa_{i}^2  +   \sum\limits_{i=1}^N \left\{ \lambda\, (\kappa_i^2 - m^2)^2
+ \frac{d}{2} \,  \kappa_i^2  \tau_i^2 \right.
\]
\begin{equation}
\left.  - \ b \kappa_i^2 \tau_i  - a  \tau_i + \frac{c}{2} \tau_i^2\right\}  
 + \mathcal O( \Delta \kappa_i^4)
\label{eq:A_energy}
\end{equation}
Here ($\lambda,m,a,b,c,d$) are parameters. For a given PDB protein structure these parameters are
determined by training a minimum energy configuration of (\ref{eq:A_energy}) to model the PDB backbone. 
There is a program {\it Propro} that can be used to train the parameters in (\ref{eq:A_energy}) 
so that the soliton profile models a given PDB structure. The program can be used on-line, it can be found at
\begin{equation}
{\tt http://www.folding-protein.org}
\label{propro}
\end{equation}

We recognise
in (\ref{eq:A_energy}) a deformation of the Hamiltonian that defines the discrete 
nonlinear Schr\"odinger (DNLS) equation ~\cite{Chernodub-2010,Molkenthin-2011}. The first row  coincides with a {\it naive} discretisation of the continuum nonlinear Schr\"odinger
equation. The fourth term ($b$) is the conserved momentum in the DNLS model,  the fifth ($a$) 
term  is the Chern-Simons term, and the sixth ($c$) term is the Proca mass. Note that both momentum and Chern-Simons are 
chiral. We refer to \cite{hu_2013,ioannidou_2014} for detailed analysis.

\subsection{Validation of mean field approach} 

In the case of proteins we validate (\ref{eq:A_energy}) qualitatively 
with the following line of arguments:  
According to ~\cite{privalov_1979,privalov_1989,Shaknovich_1989}
the  folding of a protein is a "cooperative'' process that resembles a first order phase transition. 
Indeed, the DNLS equation supports solitons which are the paradigm  cooperative organisers  in numerous physical scenarios.
A soliton emerges as a solution of the  variational equations that coincide with the extrema of (\ref{eq:A_energy}).
For this we first eliminate the torsion angles 
using the equation
\begin{equation}
\tau_i[\kappa] \ = \ \frac{a+b\kappa_i^2}{c+d\kappa_i^2}
\label{tau}
\end{equation} 
For bond angles we then  obtain
\begin{equation}
\kappa_{i+1} = 2\kappa_i - \kappa_{i-1} + \frac{dV[\kappa]}{d\kappa_i^2}  
 \label{kappaeq}
\end{equation}
where
\[
V[\kappa] = - \left( \frac{bc-ad}{d}\right) \, \frac{1}{c+d\kappa^2}  - \left( \frac{b^2 + 8 \lambda m^2}{2b}\right) \kappa^2 + \lambda \kappa^4
\]
The difference equation (\ref{kappaeq}) can be solved iteratively, for example using the algorithm in  \cite{Molkenthin-2011}; 
a soliton solution models a super-secondary protein structure such as a helix-loop-helix motif and the loop corresponds to the soliton proper. The parameter $m$ is the main regulator of the secondary structure, its value specifies 
whether we
have an $\alpha$-helix, a $\beta$-strand, or some other kind of regular pattern.  Details can be found in  \cite{Danielsson-2010,Chernodub-2010,Molkenthin-2011,Krokhotin-2012a,Krokhotin-2012b,krokhotin-2014,Sieradzan-2014,fadde,les-houches}.

In order to reveal a relation between  (\ref{eq:A_energy}) and the structure of a first order phase transition, 
we remind  that in the case of a protein the  bond 
angles are rigid and the torsion angles are flexible. In particular,
the variations of $\kappa_i$ along the backbone are  small and cover only a portion of the allowable 
range of $\kappa$ as shown in Figures \ref{fig-2} and \ref{fig-3}. 
Thus, over sufficiently large
distance scales we may try and proceed self-consistently, to ignore  fluctuations and use the mean value 
$\kappa_i \sim \kappa$. We can then solve for this  mean value $\kappa$ in terms of the mean value of torsion angles 
$\tau_i \sim \tau$. From (\ref{eq:A_energy})  
\begin{equation}
\frac{\delta E}{\delta \kappa} = 0 \ \Rightarrow \ \kappa^2 \ = \ m^2 + \frac{b}{2\lambda} \tau - \frac{d}{4\lambda} \tau^2 \ 
\label{mftk}
\end{equation}
We substitute this into the equation that determines the extrema of (\ref{eq:A_energy}) {\it w.r.t.} variations in $\tau$,
\[
\frac{\delta E}{\delta \tau} = 0 \ \Rightarrow 
\]
\begin{equation}
\frac{d^2}{4\lambda} \tau^3 -  \frac{3 bd}{4\lambda} \tau^2 + \left( \frac{b^2}{2\lambda} - dm^2 - c\right)
\tau + \left( a + bm^2\right)  = 0 
\label{mftt}
\end{equation}
This equation coincides with the variational equation that specifies the extrema of the following 
free energy
\begin{equation}
 \frac{d^2}{16 \lambda} \tau^4 - \frac{db}{4\lambda} \tau^3 + \left( \! \frac{b^2}{4\lambda} - \frac{dm^2}{2} - \frac{c}{2}\right) \tau^2 + \left( a+bm^2\! \right) \tau
\label{mftF}
\end{equation}
This has the canonical form of the Landau - De Gennes free energy for a first order phase transition ~\cite{DeGennes-1995}, thus
completing 
a {\it naive} qualitative validation of (\ref{eq:A_energy}) along the arguments in  ~\cite{privalov_1979,privalov_1989,Shaknovich_1989}. 

There are entropic corrections  that are important
in the case of proteins. The evaluation of entropic corrections and the effect of fluctuations more generally, 
proceeds in the usual manner of Landau-Ginsburg-Wilson theory \cite{Goldenfeld-1992}; here the Glauber 
algorithm accounts for entropic corrections.

\subsection{Long distance interactions in mean field theory}
\label{LJi}

A protein chain is subject to long distance interactions which in a MD approach are modeled by hydrogen bond, Lennart-Jones, 
Coulomb and other force fields. In the present mean field theory 
these interactions are presumed to be taken into account  through
the nonlinear terms in the free energy, except for the short distance Pauli repulsion which needs to be introduced explicitely.
In the leading order it suffices to proceed as follows: 
A statistical analysis of PDB structures shows that two C$\alpha$ atoms located at sites $i$ and $k$ respectively
that are {\it not} nearest neighbours along the backbone, obey the constraint 
\begin{equation}
| \mathbf r_i - \mathbf r_k | > 3.8 \ {\mathrm \AA} \ \ \ {\rm for} \ \ \ |i-k| \geq 2
\label{selfavo}
\end{equation}
We impose this constraint as a rigid acceptance criterion in the  Monte Carlo algorithm.  
More elaborate forms of short and long distance interactions could also be introduced, and 
we refer to \cite{Sinelnikova-2015}  for analysis.

\subsection{Glauber algorithm and fluctuations in mean field approach}
\label{Ga}

Arrhenius' law  states that the reaction
rate $r$ depends exponentially on the ratio of activation energy $H$ 
and the physical temperature factor $k_B \theta$,
\[
r  \ \propto  \ \exp\{ - \frac{H}{k_B \theta}\}
\]
Here $k_B$ the Boltzmann 
constant and $\theta$ is the temperature measured in  Kelvin. 
On the other hand, Glauber dynamics assumes that the transition probability from a state $a$ to another 
state $b$  has the form
\begin{equation}
\mathcal P(a \to b) \ = \ \frac{1}{ 1 + e^{\Delta E_{ba} /T} }  
\label{glauber}
\end{equation}
where $\Delta E_{ba} = E_b - E_a$ is the activation energy, and in the mean field approach
we compute it from  (\ref{eq:A_energy}).  The parameter $T$ is the Monte Carlo temperature 
factor. In general it does not coincide with the  physical temperature factor $k_B \theta$,   
but the two can be related by methods of renormalisation group \cite{Krokhotin-2013c}; here we
do not need an explicit relation, ultimately the molecular dynamics step in our algorithm will determine the
temperature.

A small two-state protein often folds in line with
Arrhenius' law \cite{Scalley-1997} and for  a simple spin chain the Glauber 
algorithm reproduces Arrhenius law. Furthermore, a protein backbone with its side-chains has
a structure that resembles a spin chain \cite{krokhotin-2014}. Thus we proceed
by assuming that Glauber dynamics is a good leading order approximation to  
model aspects of protein dynamics.

\subsection{Side chains in mean field approach}
\label{sub-R}

The mean field theory (\ref{eq:A_energy}) builds on the C$\alpha$ coordinates.  There is no direct information on the
side chain coordinates,  their effects are accounted for implicitly  by the  interactions in (\ref{eq:A_energy}). 
Once we have found a minimal energy C$\alpha$ structure, we can reconstruct an {\it Ansatz}  for the 
all atom structure using  side chain libraries; here we use  {\it Pulchra} \cite{Rotkiewicz-2008}. 
We can then employ  MD to refine the ensuing side chain structure if need be \cite{Shaw-2012}. Accordingly we proceed as follows: 
Once we have constructed a set of minimal energy C$\alpha$ structures using Glauber dynamics, we screen it 
using  {\it Pulchra} and proceed only with those C$\alpha$  structures
that are void of all atom steric clashes. For this we demand that the distance between any pair of 
atoms that are {\it not} covalently bonded,  is larger than a pre-determined 
cut-off distance $R_0$.  The covalent bond distance between C, N and O atoms is at most around
$\sim  1.54$ \AA, thus we adopt  the following global value 
\begin{equation}
R_0 = 1.6 \ {\rm \AA}
\label{R0}
\end{equation}
We only proceed to our MD stability analysis with such mean field structures that pass this screening.

\subsection{Molecular dynamics}

We use the molecular dynamics package GROMACS 4.6.3 \cite{Hess-2008}. We analyse in detail 50 ns 
long trajectories, with initial configurations that we specify in the sequel. We use the  Gromos53a6 force field,
with a time step of 2 fs. We motivate our choice of force field by the analysis in \cite{Dai-2016b}. 
We use periodic boundary conditions, with 0.9 nm cut off for long distance interactions. The box is rectangular,
with a distance of 2.0 nm between the protein and the box walls; we adjust the box size depending
on the initial configuration. We use  salt concentration of 0.15 mol/l and temperature 290 K, 
supported with Berendsen thermostat in the equilibration phase and with v-rescale thermostat in the production run. 
The pressure is kept constant with Berendsen barostat, and changed to Parrinello-Rahman in the
production run. The changes ensure that we generate a proper canonical
ensemble. We record the coordinates every 2 ps, which gives us  2500 frames for each poduction run.

\section{Results}

\subsection{Myc as a Multisoliton}

Our starting point is the crystallographic PDB structure with  code 1NKP  \cite{Nair-2003} shown in Figure \ref{fig-4}. 
It describes a DNA bound heterodimer of Myc and Max in a base-helix-loop-helix leucine zipper conformation, 
with 1.8 \AA ~ resolution. We use chain A where the sites with PDB index 897-984 correspond
to Myc, its major features are summarised in Table \ref{table-1}.
\begin{table}[h]
\renewcommand\arraystretch{1.5}
\centering
\begin{tabular}{clc}
\Xhline{2pt}
Structure &                           ~~~    Residues &                                       Number of residues\\
\Xhline{1pt}     
\hline
base&                                                      GLY897-ARG914&                                               18\\
\hline
helix&                                                      ASN915-ILE928&                                               14\\
\hline
loop&                                                    PRO929-PRO938&                                               10\\
\hline
helix&                                                   LYS939-CYS984&                                               46\\
\Xhline{2pt}
\end{tabular}
\caption{Structure assignment of Myc in PDB entry 1NKP chain A.}
\label{table-1}
\end{table}

In Figures \ref{fig-5} top and bottom we show the bond and torsion angles of Myc in 1NKP, respectively. 
In these Figures we follow \cite{Hu-2011} and extend 
the range of bond angle into $\kappa \in [-\pi,\pi)$.  Thus there is a two-fold covering of the geometry by the
bond angles, which we compensate for with a discrete local $\mathbb Z_2$
symmetry. We employ this symmetry to identify the soliton content, which is visible in Figure {\ref{fig-5} (Top): We identify six
different individual solitons. Four of the solitons form the
loop region of Myc.  There is one soliton along the leucine zipper, at the location of the 
turn around residue 954 where we observe a jump in the torsion angle in Figure \ref{fig-5}. There is one soliton at
the C terminal of Myc.
%
%
%
%
%
%
%
%
%
%
%
%
%
\begin{figure}[h]         
\centering            
\vspace{0.2cm}  \resizebox{8 cm}{!}{\includegraphics[]{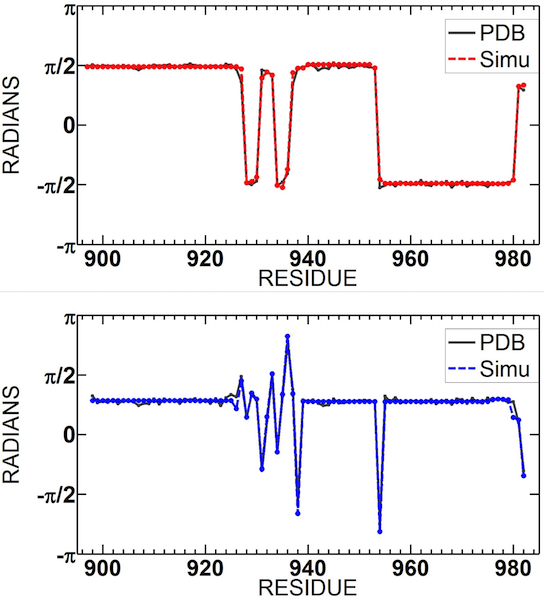}}
\caption {\small  {\it Color online:} (Top:) The bond $\kappa$ angle spectrum of 1NKP after we introduce
the $\mathbb Z_2$ transformation that identifies the soliton structure (grey), together with the
corresponding spectrum of the multisoliton structure (red). (Bottom:)  The torsion $\tau$ 
angle spectrum of 1NKP after the $\mathbb Z_2$ transformation that identifies the 
soliton structure in top figure (grey), together with the
corresponding spectrum of the multisoliton structure (blue).
}   
\label{fig-5}    
\end{figure}
%
%
%
%
%
%
%
%
In Figures \ref{fig-5} we also compare the bond and torsion angle values between Myc in 1NKP and its multisoliton.
We use the program {\it Propro}  that can be found from  (\ref{propro}) to construct the multisoliton profile, the 
parameter values that we find for the  Landau free energy are given in Table \ref{table-2}; we have rounded the numbers to
four digits, higher precision is easily obtained using  (\ref{propro}). 

\begin{table}[!htb]
\renewcommand\arraystretch{1.5}
\centering
\begin{tabular}{|c|c|c|c|}
\hline
parameter &	soliton-1 &    	soliton-2 &       soliton-3  \\ 
\hline
b&	         ~  8.731 \!\!\! e-11~ &	~ 1.588  \!\!\! e-8~&  ~    6.486 \!\!\! e-10~ \\
\hline
d&              1.268 \!\!\!  e-7&     7.637 \!\!\!  e-8&       5.832 \!\!\!  e-8  \\
\hline
e&              1.004 \!\!\! e-11&     8.77 \!\!\!  e-10&      1.888 \!\!\!  e-10 \\
\hline
q&              -5.944 \!\!\!  e-8&     -6.827 \!\!\!  e-7&       -5.755 \!\!\!  e-9 \\
\hline
c1&             5.459&          2.603&          4.459 \\
\hline
c2&             2.318&          2.252&          4.13 \\
\hline
m1&             1.539&          1.494&          1.405 \\
\hline
m2&             1.651&         1.404&         1.655 \\
\hline
\end{tabular}

\vspace{0.2cm}

\begin{tabular}{|c|c|c|c|}
\hline
parameter &	soliton-4 &    	soliton-5 &       soliton-6  \\ 
\hline
b&	           ~3.414  \!\!\!  e-10 ~&~      4.4967 \!\!\! e-9 ~&~    1.148 \!\!\! e-9   ~\\
\hline
d&              4.865 \!\!\! e-8 &        6.943 \!\!\!  e-10 &  3.402 \!\!\! e-8      \\
\hline
e&              1.501 \!\!\!  e-10 &       7.612 \!\!\! e-15 &   1.353 \!\!\!  e-10   \\
\hline
q&              -5.331 \!\!\!  e-8 &      -2.906 \!\!\!  e-7 &   - 7.716  \!\!\!  e-8    \\
\hline
c1&             0.887&          3.225 &   2.872        \\
\hline
c2&             2.449&          2.995 &     18.223      \\
\hline
m1&             1.533&          1.595 &   1.54        \\
\hline
m2&             1.5 &          1.54 &   1.049       \\
\hline
\end{tabular}
\caption{Parameters for each soliton.}
\label{table-2}
\end{table}

In Figure \ref{fig-6} we interlace the original PDB structure of Myc with the multisoliton profile. The root-mean-square distance
(RMSD) between the two is 0.98 \AA. 
%
%
%
%
%
%
%
%
%
\begin{figure}[h]         
\centering            
\vspace{0.2cm}  \resizebox{8 cm}{!}{\includegraphics[]{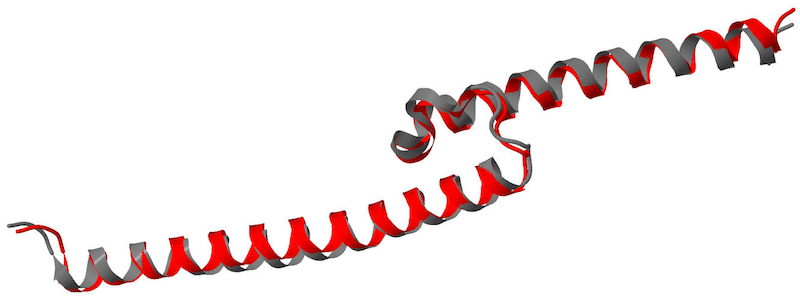}}
\caption {\small  {\it Color online:} Comparison between the crystallographic Myc (grey) and its multisoliton (red).
}   
\label{fig-6}    
\end{figure}
%
%
%
%
In Figure \ref{fig-7}
%
%
%
%
%
%
%
%
%
%
%
%
%
\begin{figure}[h]         
\centering            
\vspace{0.2cm}  \resizebox{8 cm}{!}{\includegraphics[]{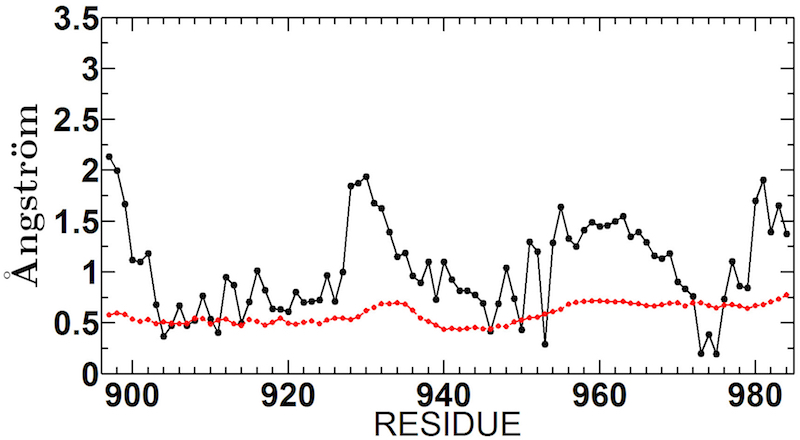}}
\caption {\small  {\it Color online:} Residue-wise distance of the C$\alpha$ atoms in the multisoliton from their crystallographic 
coordinates. The red line shows the Debye-Waller one-$\sigma$ fluctuation distance in the PDB structure. }   
\label{fig-7}    
\end{figure}
%
%
%
%
we compare the residue-wise distance of the C$\alpha$ atoms in the multisoliton from those in the crystallographic structure. We
also show an estimate of the  one standard deviation  error in the crystallographic coordinates, which we compute from the
B-factors using the Debye-Waller relation.

\subsection{Glauber attractor and minima of Landau free energy}

We subject the multisoliton to successive 
heating and cooling simulations using the Glauber algorithm. Our 
goal is to identify the low temperature Glauber attractor. This we define as the set
of all the structures towards which the Glauber algorithm converges in 
the limit of vanishing ambient temperature. {\it A priori}, the Glauber attractor should coincide with the landscape of local 
minima of the Landau free energy.

As  we increase the ambient temperature,  the multisoliton structure starts to thermally fluctuate. At sufficiently high temperatures 
the structure can cross over energy barriers that surround the initial multisoliton profile. If the multisoliton is not stable, 
the structure can be expected to start drifting  away from its
vicinity. 
When the ambient temperature subsequently decreases, the structure becomes attracted towards 
a local minimum of the  free energy.  In the case of a protein such as
myoglobin that has an essentially unique and stable fold,  the final conformation coincides with the initial 
multisoliton. However,  if the protein is intrinsically unstructured there are in general several
local minima in the  free energy landscape to which it can become attracted; the final conformation does not need to be unique. In fact, it
can be quite different from the initial one. The Glauber attractor can have an elaborated topography.  
When we repeat the heating and cooling cycle sufficiently many times,  with sufficiently high temperature variations,
we expect to eventually resolve for the Glauber attractor.

In the case of Myc we deduce that $T_H=10^{-9}$ and $T_L= 10^{-14}$ can be considered   
representative for the high and low temperature factor values, that we use  in  the 
Glauber algorithm (\ref{glauber}). 
Similarly, we conclude that  $50 \times 10^6$  is a representative number of Monte Carlo steps for 
a heating and cooling cycle.  In a single cycle we first take $5\times10^6$ steps at  the low temperature factor value $T_L$,
to fully thermalise the system.  We then increase $T$ during $10\times10^6$ steps linearly  on a logarithmic temperature scale, to
the high temperature factor value $T_H$, and   thermalise the system at $T_H$ 
during $20\times10^6$ steps. We conclude the cycle by lowering $T$ back to $T_L$, by reversing the 
heating process. We repeat the cycle until we are confident that we have identified the Glauber attractor.    
We have performed several repeated heating and cooling simulations,  with 
the number of cycles varying between 2500 and 5000.
Accordingly we have very good statistics.  There are  two production runs that we analyse  in detail. In one we use the full length
Myc structure while in  the other we exclude the residues near the flexible N and C terminals, and 
simulate only the segment between amino acids 901-979. We find the same  
Glauber attractor in both cases.  We always perform all  MD simulations  with 
the full chain length.
 
The Figures \ref{fig-8}  show the statistical evolution of RMSD and the radius of gyration $R_g$ that we 
obtain in all our heating and cooling cycles. 
%
%
%
%
%
%
%
%
%
%
%
%
%
\begin{figure}[h]         
\centering            
\vspace{0.2cm}  \resizebox{8 cm}{!}{\includegraphics[]{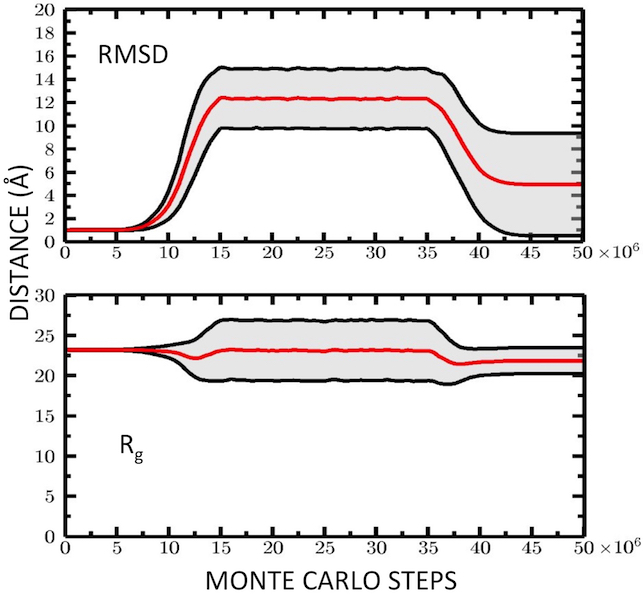}}
\caption {\small  {\it Color online:} (Top:) Evolution of RMSD distance from the crystallographic Myc structure during the heating and cooling cycle. (Bottom:) Evolution of radius of gyration $R_g$ during the heating and cooling cycle. The red line denotes the
average value over 5175 simulations, and the grey band determines the 
one-$\sigma$ deviation from the average value.}   
\label{fig-8}    
\end{figure}
%
%
%
%
There is in average around 5 \AA ngstr\"om RMSD distance  between the initial 
crystallographic structure and the final structure. This distance is larger than the $\sim$4.2 \AA
~one standard deviation spread that we observe in the average value of the RMSD distance from the crystallographic Myc. 
Thus the Glauber attractor is  degenerate,  there does not appear to be a single folded state. 
In particular, the initial multisoliton is  not stable. Such a degeneracy can 
be expected, in the case of an intrinsically unstructured protein.

In Figure \ref{fig-9} we present the Glauber attractor on the 
RMSD {\it vs.} $R_g$ plane. In this Figure we identify five disjoint clusters for our 
future analysis. One of the clusters (number 5) corresponds to structures that return to the vicinity
of the initial multisoliton of the crystallographic Myc. We observe that the full Glauber attractor is tightly located around
the line
\begin{equation}
{\rm RMSD} \ \approx - 2.4 \, R_g \ +  \ 56.9 \ \ \ \ ({\rm \AA})
\label{center}
\end{equation}
Figure \ref{fig-9}  shows that the initial multisoliton  has a tendency to  collapse towards 
spatially more compact structures.   
%
%
%
%
%
%
%
%
%
%
%
%
%
\begin{figure}[h]         
\centering            
\vspace{0.2cm}  \resizebox{8 cm}{!}{\includegraphics[]{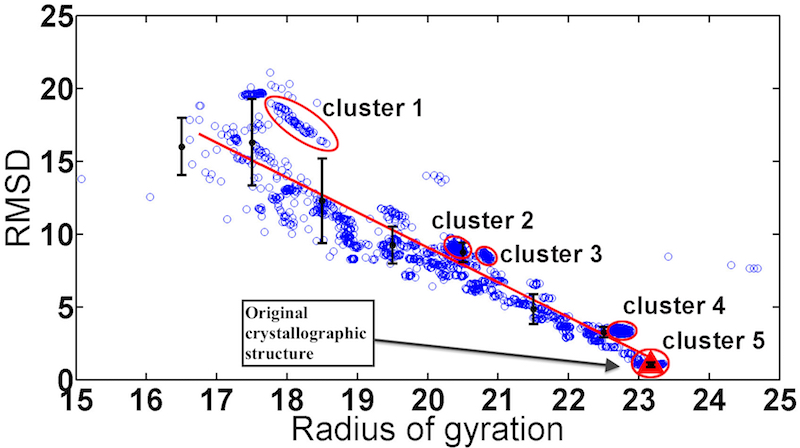}}
\caption {\small  {\it Color online:} The low temperature Glauber attractor
on the $R_g$ {\it vs.} RMSD plane. Five different clusters are identified,
and the initial multisoliton is marked with a red triangle in cluster 5. The error-bars denote 
one standard deviation distance around the average value, that determines the average line (\ref{center}).
}   
\label{fig-9}    
\end{figure}
%
%
%
%

Figure \ref{fig-10} shows the Landau free energy landscape of Glauber attractor as a function of $R_g$ and RMSD. 
%
%
%
%
%
%
%
%
%
%
%
%
%
\begin{figure}[h]         
\centering            
\vspace{0.2cm}  \resizebox{8 cm}{!}{\includegraphics[]{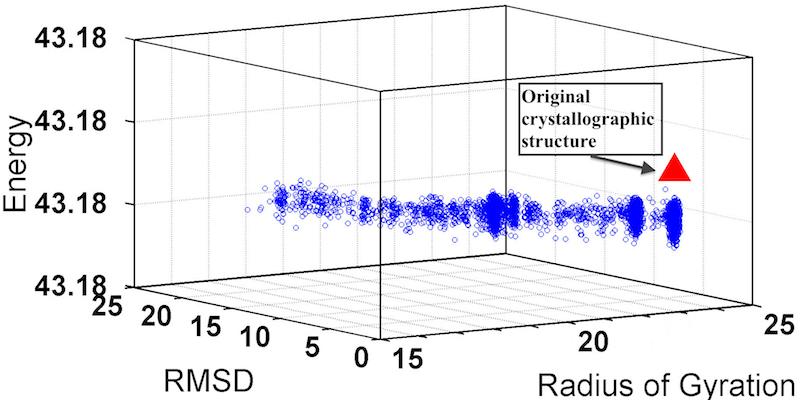}}
\caption {\small  {\it Color online:} (Top:) The energy landscape of Glauber attractor, in terms of $R_g$ {\it vs.} RMSD of the final structures. The multisoliton is marked with a red triangle. Note that the energy of the attractor is highly degenerate, free energy differences are minor.
The RMSD and Radius of Gyration are measured in \AA ngstr\"om, energy unit is defined by the overall normalisation of (\ref{eq:A_energy}).
}   
\label{fig-10}    
\end{figure}
%
%
%
%
The initial 
multisoliton is marked by red triangle. It is unstable, its Landau free energy is above that
of Glauber attractor states.   We note that the free energy is highly degenerate, to the extent that 
he clusters of Figure \ref{fig-9} appear as space filling point sets, prolated along the energy axis on 
the scale that we use in the Figure. 
A space filling ground state degeneracy is in line with the expected intrinsically unstructured character of Myc.
Thermal fluctuations move the structure around, making it to hop between the different low temperature states 
near the local energy minima.

In Table \ref{table-3} we summarise the main characteristics of the clusters that we
identify in Figures \ref{fig-9} and \ref{fig-10}. 
\begin{table}[h]
\renewcommand\arraystretch{1.5}
\centering
\begin{tabular}{clclc}
\Xhline{2pt}
Cluster &                          ~ min &              max      \\
\Xhline{1pt}     
\hline
1 &                                 ~    16.2 ~ ~ &                  ~                18.7 ~  \\
\hline
2 &                                  ~    8.7  ~ ~ &                   ~          9.2 ~   \\
\hline
3 &                                  ~    8.3  ~ ~ &                  ~        8.8   ~              \\
\hline
4 &                                  ~      3.1 ~ ~ &                 ~     3.4          ~   \\
\hline
5   &                           ~       0.9 ~ ~ &             ~   1.2 ~  \\
\Xhline{2pt}
\end{tabular}
\caption{Minimimum and maximum RMSD distance of each cluster from the initial Myc structure in 1NKP.}
\label{table-3}
\end{table}

\subsection{Molecular dynamics analysis}

We use MD  to analyse the {\it local} spatial and temporal stability of clusters in the Glauber attractor.
We perform the simulations at relatively low, near room temperature value 290 K. 
Higher temperature entails larger amplitude thermal fluctuations and it becomes
difficult to deduce the level of local cluster stability in the background of large amplitude thermal motions.

The Landau free energy engages only the C$\alpha$ backbone, thus it can support backbone structures
with steric clashes in all atom structures.   We start by screening out such structures, that lead to steric clashes.
For this we use  {\it Pulchra} side chain reconstruction algorithm, to recover an all atom structure from
the C$\alpha$ trace.  We impose {\it stringently} the acceptance criterion  (\ref{R0}).  
The Figure \ref{fig-11} shows the overall acceptance ratio of mean field  C$\alpha$  structures as a function of the 
cut-off parameter $R_0$. 
%
%
%
%
%
%
%
%
%
%
%
%
%
\begin{figure}[h]         
\centering            
\vspace{0.2cm}  \resizebox{8 cm}{!}{\includegraphics[]{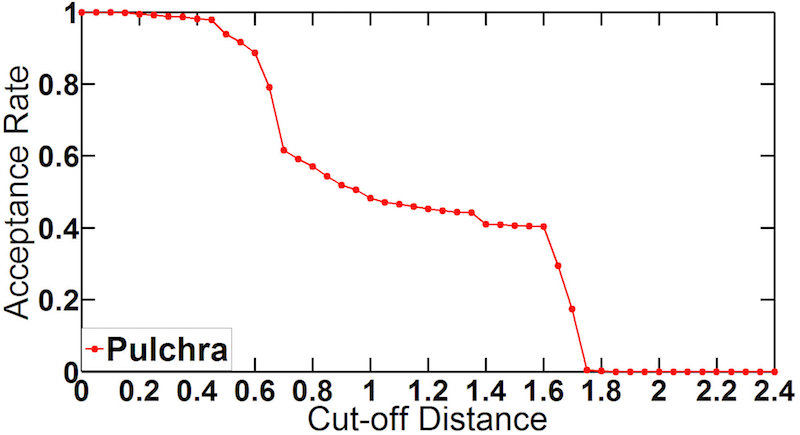}}
\caption {\small  {\it Color online:} The fraction of structures in the Glauber attractor with not a single all atom steric clash
according to {\it Pulchra}, as a function of the cut-off parameter $R_0$. Here we adopt the cut-off value $R_0 = 1.6$ \AA.
}   
\label{fig-11}    
\end{figure}
%
%
%
%
At the value (\ref{R0}) of $R_0$ around 40 per cent of {\it Pulchra} structures are sterically fully consistent.
In particular, each of the five clusters have  representatives with sterically acceptable all atom structures.   
But we note a sharp drop in the number of accepted structures when $R_0$ increases beyond the value (\ref{R0}).

We (randomly) select  one all atom structure from each
of the five clusters for  MD stability analysis.  We find  that only  cluster 1  
appears MD stable. In all other clusters, the initial structure drifts systematically away from the cluster 
under MD  time evolution.  
We describe the examples in clusters 1,4 ad 5 in more detail:  
In Figure \ref{fig-12} (Top) we show the evolution of RMSD and in Figure \ref{fig-12} (Bottom) 
we show the evolution of $R_g$ during our MD simulations.  
%
%
%
%
%
%
%
%
%
%
%
%
%
\begin{figure}[h]         
\centering            
\vspace{0.2cm}  \resizebox{8 cm}{!}{\includegraphics[]{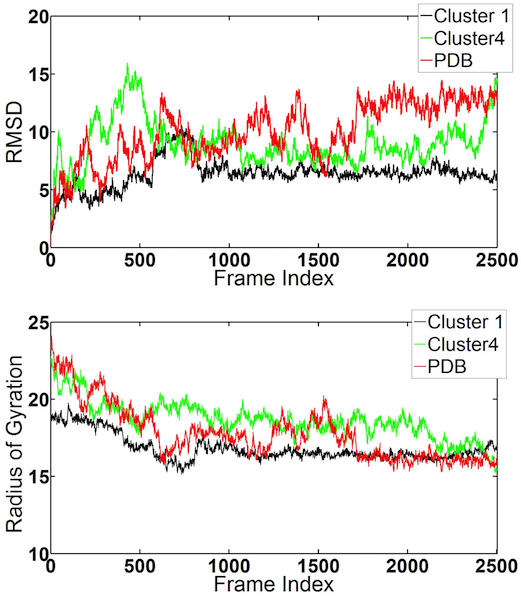}}
\caption {\small  {\it Color online:} (Top:) The evolution of RMSD 
in the three clusters 1,4 and the PDB structure 5   under our MD simulations. (Bottom:) The evolution of $R_g$ 
in the three clusters 1,4 and 5 under our MD simulations. The RMSD is computed from the initial structure 
of the simulation, chosen randomly in cluster 1 and 4, and as the multisoliton in cluster 5.
}   
\label{fig-12}    
\end{figure}
%
%
%
%
%
%
%
%
We note how in both cluster 4 and 5 the initial structure drifts away from the cluster. In cluster 1
the structure also initially moves away, but then the values of RMSD and $R_g$ quickly stabilise:
Qualitatively, the evolution in cluster 1 is different from the other two. There is an apparent initial 
relaxation of the tension in the {\it Pulchra} side chain assignment, with corresponding adjustment of the
backbone during the first $\sim$15 nanoseconds. This is followed by a stabilisation.  In Figure \ref{fig-13} we compare the evolution 
trajectory for cluster 1, with the cluster 5 of PDB structure. The cluster 1 converges towards
a region which is close to the original cluster, while cluster 5 systematically 
drifts away from the initial position. The MD trajectories of cluster 1 and 5 are quite
different. 
%
%
%
%
%
%
%
%
%
\begin{figure}[h]         
\centering            
\vspace{0.2cm}  \resizebox{8 cm}{!}{\includegraphics[]{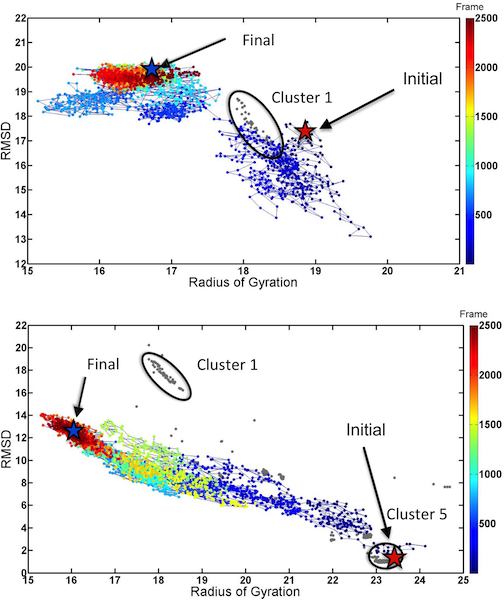}}
\caption {\small  {\it Color online:} (Top:) The evolution of cluster 1 under MD simulation.
(Bottom:) The evolution of cluster 5 (PDB structure) under MD simulation. Color coding for time evolution is same
in both Figures, but note the difference in scales.  
}   
\label{fig-13}    
\end{figure}
%
%
%
%
%
%
%
%


\subsection{Myc at room temperature}

The mean field structures in the cluster 1 have the shape of a hairpin. In  top Figure \ref{fig-14} we show  
a generic structure,  the one that we use as the initial configuration
in our  MD simulation. The two $\alpha$-helical segments of Figure \ref{fig-6}
have become almost parallel, and quite close to each other. 
%
%
%
%
%
%
%
%
%
%
%
%
%
\begin{figure}[h]         
\centering            
\vspace{0.2cm}  \resizebox{8 cm}{!}{\includegraphics[]{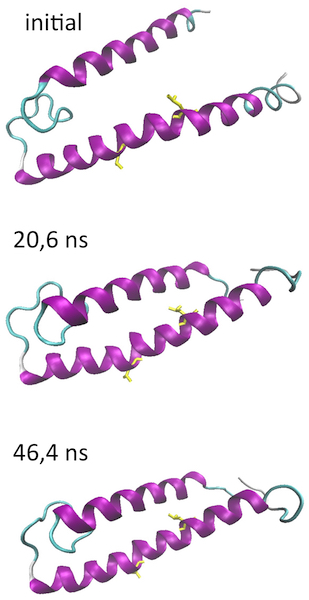}}
\caption {\small  {\it Color online:} (Top:) The initial structure  from cluster 1, in the all atom MD simulation. (Middle 
and Bottom:) Two representative snapshot structures from around halfway and near end of the MD simulation.
In the Figures we have identified two residues Leu-951 and Leu 960. 
}   
\label{fig-14}    
\end{figure}
%
%
%
%
In the course of the MD time evolution the hairpin continues to emerge, but intermittently:
The hairpin  shown
in the middle and bottom Figures \ref{fig-14}  repeats itself several times during the time evolution; 
the middle Figure is taken near the halfway point of the MD simulation 
and the bottom Figure is taken close to its end. 
The two hairpins are almost identical, and very similar to  the (generic) mean field cluster 1 structure that we show in the 
top Figure.

Besides the hairpin of Figure \ref{fig-14} we identify another structure that appears repeatedly in our MD simulation.
We show it both in the top (9.74 ns) and bottom (18.52 ns) snapshots of the motion we outline in Figures \ref{fig-15}. 
%
%
%
%
%
%
%
%
%
\begin{figure}[h!]         
\centering            
\vspace{0.2cm}  \resizebox{6.0cm}{!}{\includegraphics[]{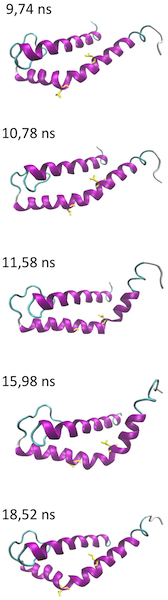}}
\caption {\small  {\it Color online:} The MD evolution of the C$\alpha$ backbone between frames 443-786 of Figure \ref{fig-13}.
}   
\label{fig-15}    
\end{figure}
%
%
%
%
%
%
%
%
%
%
This structure is essentially the hairpin of Figures \ref{fig-14}, but with a turn near  the middle of one of the two parallel 
helices. The turn is located 
right after Leu-951,  in the proximity of the turn 
that we observed previously in the crystallographic structure; see
Figure \ref{fig-5}.  

We deduce that the MD trajectory 
is akin a dynamical two-state system, with oscillatory motion between the hairpin structure of Figures \ref{fig-14}
and the turn-in-helix hairpin structure that we show in the top and bottom snapshots of Figure \ref{fig-15}. 
In Figure \ref{fig-16} we confirm this, in this Figure we show the results from a secondary
structure analysis of the entire MD trajectory, starting from cluster 1. 
%
%
%
%
%
%
%
%
%
\begin{figure}[h!]         
\centering            
\vspace{0.2cm}  \resizebox{8.0cm}{!}{\includegraphics[]{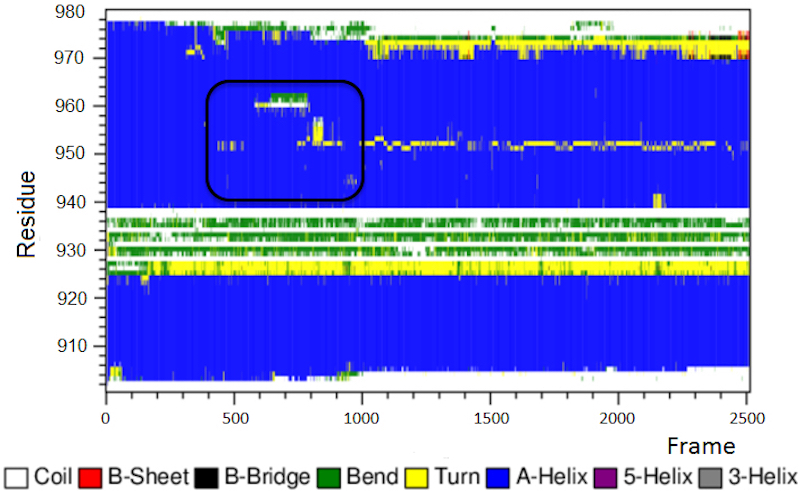}}
\caption {\small  {\it Color online:} The {\tt do.dssp} secondary
structure analysis of the MD trajectory. The apparent Davydov soliton in Figures \ref{fig-15} is identified.
%
%
%
}   
\label{fig-16}    
\end{figure}
%
%
%
%
%
%
%
%
%
%
The secondary structure profile is remarkably stable, except for  the oscillatory two state dynamics between 
the  hairpin of Figure \ref{fig-14} and the hairpin with turn-in-helix. 
The oscillations between the two states start after an initial hairpin stabilisation period of 
around 8 ns. 

In Figure \ref{fig-17} (Top) we 
show the evolution of an angle which is formed between the 
vector pointing from the C$\alpha$ to C$\beta$ at site Leu-951, and the corresponding vector at site Ala-955.
%
%
%
%
%
%
%
%
%
\begin{figure}[h!]         
\centering            
\vspace{0.2cm}  \resizebox{8.0cm}{!}{\includegraphics[]{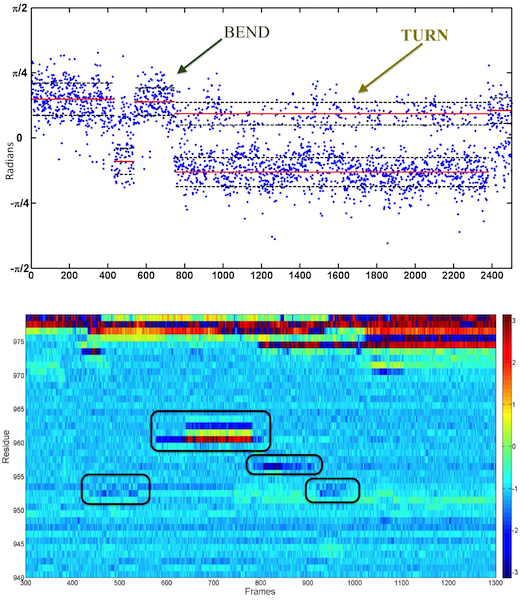}}
\caption {\small  {\it Color online:} (Top:) The angle between vector pointing from C$\alpha$ to C$\beta$ at Leu-951, 
and the vector pointing from C$\alpha$ to C$\beta$ at Ala-955. (Bottom:) 
The values of the side chain $\eta$-angle \cite{Dai-2016b} during the frames 200-1300 of the MD simulation of cluster 1.
We have encircled the portion that relates to the event in Figure \ref{fig-15}, also identified in Figure \ref{fig-16}. 
}   
\label{fig-17}    
\end{figure}
%
%
%
%
%
%
%
%
%
%
The dynamical two-state oscillatory character of the trajectory is apparent in this Figure: After the initial stabilisation, the 
angle between the two vectors jumps between two different values,
corresponding to the helix and to the turn-in-helix structures.  Such oscillatory behaviour between multiple  
different, energetically degenerate structures  
has been previously identified in the case of intrinsically unstructured proteins \cite{Dai-2016}. 
Energy degeneracy with several different conformational states, separated by small energy barriers,
appears to be symptomatic for these proteins.

In the bottom Figure \ref{fig-17} we show the evolution of the side-chain $\eta$-angles along the entire chain during the 
MD time evolution. The concept  of the $\eta$-angle has been introduced and analysed  in  \cite{Dai-2016b}. The
angle measures difference in the direction between neighboring C$\alpha$-C$\beta$ vectors,   it  characterises the local
twisting of the side chain assignment along the backbone.
The focus in the Figure is in the time period between frames 200  and 1300, corresponding to time segment  4.0 - 26.0 ns.
In the Figure we identify  one relatively long-lived bend intermediate, with the bend  located near Leu-960. 
This intermediate originates from the
turn at Leu-951, and we summarise  its emergence and evolution in Figures \ref{fig-15}; see also Figure \ref{fig-16} and Figure 
\ref{fig-17} top where the intermediate is identified.  The bend formation starts with the 
turn first appearing near Leu-951, as shown in the  9.74 ns snapshot of Figure \ref{fig-15}. This turn 
propagates  to the vicinity of Leu-960, as shown in snapshots at 10.78 ns and 11.58 ns.  It stays there as a bend (according
{\tt do.dssp}) for several nanoseconds, 
and then translates back towards Leu-951 (snapshot at 15.98 ns) where it stops and forms a turn  (according
{\tt do.dssp}) as shown in the snapshot at
18,52 ns. Finally, the turn dissolves and the structure returns to the hairpin conformation of Figures \ref{fig-14}.  
The entire oscillatory event lasts around 9 nanoseconds,  and it is clearly identifiable in our simulations. 

We propose that the event we summarise in Figures \ref{fig-15}, \ref{fig-17} (bottom)  corresponds to a formation and 
propagation of Davydov's Amide-I soliton along $\alpha$-helix: Apparently the hydrogen bonds that stabilise 
the $\alpha$-helix occasionally break,   causing the formation of a turn near Leu-951. This turn  is a localised quasiparticle 
akin Davydov's soliton. It propagates along the backbone to the vicinity of Leu-960, 
bounces back, and return to Leu-951 where localises and then dissolves.

\section{Summary}

We have proposed to combine an effective mean field description 
with molecular dynamics. The outcome is a multiscale algorithm that can be used to
model protein dynamics efficiently both over long time periods and with atomic level precision.
We have applied the algorithm to study properties of  Myc, which is a biomedically highly relevant oncoprotein. 
Myc has an important role in regulation of gene expression, and  a malfunctioning or  over-expressed Myc 
has been implicated in many cancers from  Burkitt's lymphoma and neuroblastomas to carcinomas of colon, 
breast and lungs. Accordingly Myc is subject to vigorous pharmaceutical and biomedical research, it is
a potentially highly important target to anti-cancer drugs. 
 
An isolated monomeric Myc is presumed to be intrinsically unstructured under physiological conditions. 
Moreover,  as a  momoner  Myc has no known biological function, it becomes biological active only in a 
heterodimer with protein Max.  The heteromerization rate of Myc and Max  should depends on the 
conformational state of an isolated monomeric Myc {\it in vivo}. Thus, the  investigation of the conformational landscape 
in the case of an isolated Myc 
should have direct biomedical relevance:  When we understand the physical properties of a monomeric Myc, we 
can  identify mechanisms to control the rate how Myc and Max  heterodimerize. 

We have found that at room temperature a monomeric Myc has a tendency to turn into a hairpin-like conformation.
We have also found that this conformation is unstable. It tends to occasionally buckle, at a specific location that we have 
identified. Moreover, we have observed that the ensuing deformation can  translate back and forth along the 
backbone, in a manner that resembles the propagation of Davydov's Amide I soliton. Accordingly the low energy 
landscape  of Myc is degenerate, there is at least a two-state structure between which Myc oscillates at room
temperature.  The oscillatory behaviour is in line with the expected character of Myc as an intrinsically unstructured protein.

\section{Acknowledgements:}
This research was supported in part by Bulgarian Science Fund (Grant DNTS-CN 01/9/2014) and 
China-Bulgaria Intergovernmental S\&T Cooperation Project at Ministry of Science and Technology of P.R. China (2014-3). AJN acknowledges support from Vetenskapsr\aa det, Carl Trygger's Stiftelse f\"or vetenskaplig forskning, 
and  Qian Ren Grant at Beijing Institute of Technology.

\end{document}